\documentstyle{article}
\begin{document}

\title{Squashed States of Light: \\ Theory and Applications to Quantum 
Spectroscopy \\ $\phantom{.}$ \\ 
{\large Running Title:} {\Large Quantum Spectroscopy with Squashed Light}} 
\author{H. M. Wiseman  \\ Centre for Laser Science, The 
Department of Physics, \\ The University of Queensland, Queensland 
4072 Australia. \footnote{E-mail: wiseman@physics.uq.edu.au.}}  
\date{}
\maketitle

\newcommand{\beq}{\begin{equation}} 
\newcommand{\eeq}{\end{equation}}
\newcommand{\bqa}{\begin{eqnarray}} 
\newcommand{\eqa}{\end{eqnarray}}
\newcommand{\erf}[1]{Eq.~(\ref{#1})}
\newcommand{\nn}{\nonumber} 
\newcommand{\dg}{^\dagger}
\newcommand{\smallfrac}[2]{\mbox{$\frac{#1}{#2}$}}
\newcommand{\bra}[1]{\langle{#1}|} 
\newcommand{\ket}[1]{|{#1}\rangle}
\newcommand{\ip}[1]{\langle{#1}\rangle}
\newcommand{\sch}{Schr\"odinger } 
\newcommand{\schs}{Schr\"odinger's }
\newcommand{\hei}{Heisenberg } 
\newcommand{\heis}{Heisenberg's }
\newcommand{\half}{\smallfrac{1}{2}} 
\newcommand{\bl}{{\bigl(}}
\newcommand{\br}{{\bigr)}} 
\newcommand{\ito}{It\^o }
\newcommand{\str}{Stratonovich } 
\newcommand{\rt}[1]{\sqrt{#1}\,}
\newcommand{\ve}{\varepsilon}

\begin{abstract}
	Using a feedback loop it is possible to reduce the fluctuations in 
	one quadrature of an in-loop field {\em without} increasing the 
	fluctuations in the other. This effect has been known for a long time, 
	and has recently been called ``squashing'' [B.C. Buchler {\em et 
	al.}, Optics Letters {\bf 24}, 259 (1999)],  
	as opposed to the ``squeezing'' of a free field 
	in which the conjugate fluctuations are increased. 
	In this paper I present a general theory of squashing, including 
	simultaneous squashing of both quadratures and simultaneous squeezing 
	and squashing.  I show that a two-level 
	atom coupled to the in-loop light feels the effect of the 
	fluctuations as calculated by the theory. In 
	the ideal limit of light squeezed in one quadrature and squashed in 
	the other, the atomic decay can be completely suppressed.
\end{abstract}

\section{Introduction}

Squeezed states of light are nonclassical \cite{Wal86}. The foremost 
consequence of this is that 
they can produce a homodyne photocurrent having a noise level below the 
shot-noise limit. The shot-noise limit is what is predicted by a 
theory in which the light is classical, with no noise,  
but the process of photo-electron emission is treated quantum-mechanically.  

There is, however, 
a simple way to produce a sub-shot-noise photocurrent 
without squeezed light:  modulating the light incident on 
the photodetector by a current originating from that very detector.
This was first observed \cite{WalJak85,MacYam86} around the same time as the 
first incontestable observation of squeezing \cite{Slu85}. 

The sub-shot noise spectrum of an in-loop photocurrent is not regarded 
as evidence for squeezing for a number of reasons. First, the 
two-time commutation relations for an in-loop field are not those of a free 
field \cite{Sha87}. This means that it is possible to reduce the 
fluctuations in the measured (amplitude) quadrature without increasing those in 
the other (phase) quadrature. Second, attempts to remove some of the supposedly 
low-noise light by a beam splitter yields only above shot-noise light, 
as verified experimentally \cite{WalJak85,YamImoMac86}.

Because of these differences, the presence of a sub-shot noise photocurrent 
spectrum for in-loop light has by and large been omitted from 
discussions of squeezing \cite{Gar91,WalMil94,Bac98}. Nevertheless, 
it has been argued \cite{Tau95} that the 
in-loop field can justifiably be called ``sub-Poissonian'' (even if 
not squeezed), because a 
perfect quantum-non-demolition (QND) intensity meter for the in-loop 
light would register 
the same sub-shot-noise statistics as the  (perfect) in-loop detector. 
 Furthermore, it was proposed in 
Ref.\cite{Tau95} 
that the apparent  in-loop noise reduction could be used to improve the
signal to noise ratio for a measurement of a modulation in 
the coupling coefficient of such a QND intensity meter.

Following on from Ref.~\cite{Tau95}, 
it has been shown that in-loop optical noise suppression 
may have other, more practical, applications.
 Buchler {\em et al.} showed \cite{Buc99} that such in-loop light can 
suppress radiation pressure noise in a gravitational wave detector 
by a factor of two. Even more strikingly, 
I recently showed \cite{Wis98b} that a two-level atom 
coupled to the in-loop field can 
exhibit linewidth narrowing exactly analogous to that produced by 
squeezed light \cite{Gar86}.

In this paper I will discuss the properties of ``squashed'' states of 
light, as the in-loop analogues of squeezed states of light are called 
in Ref.~\cite{Buc99}. Section 3 covers the general theory of squashed 
states of light, including states which are squashed in both 
quadratures and states which are simultaneously squashed and squeezed.
In Section 4, I generalize the analysis of 
Ref.~\cite{Wis98b} by considering the effects of 
these more general state of light on 
a two-level atom. But to begin, I review the properties of squeezed 
states of light in the following section.

\section{Squeezed States}

\subsection{Single-mode squeezing}

The noise in the quadratures of a single-mode light field of 
annihilation operator $a$, 
\beq
x = a+a\dg \;,\;\; y = -ia +ia\dg,
\eeq
is limited by the \hei uncertainty relation
\beq
V_{x} V_{y}  \geq \left|\half[x,y]\right|^{2}=1,
\eeq
where $V$ denotes variance.  
Squeezed states of light are states such that 
one of the quadrature variances, say $V_{x}$, is less than one 
\cite{Wal86,Cav81,Yue76}. Clearly the other quadrature variance,
say $V_{y}$, must be greater than one. If the equality is attained in 
the uncertainty relation then these are called minimum uncertainty 
squeezed states. The only other  sort of minimum 
uncertainty state is the coherent state with $V_{x}=V_{y}=1$. 

\subsection{Continuum Squeezing}

Single-mode squeezing was generalized early to multimode squeezing 
\cite{CavSch85}. In this work I wish to consider the limit of an 
infinite number of modes: the  
electromagnetic continuum. Considering polarized light propagating in 
one direction, only a single real-valued index is needed for the 
modes, and we take that to be the mode frequency $\omega = k$ (using 
units such that the speed of light is one). Then the continuum field 
operators $b(k)$ obey
\beq
[b(k),b\dg(k')] = \delta(k-k').
\eeq

For light restricted in frequency to a relatively narrow bandwidth 
$B$ around a carrier frequency $\Omega$ it is possible 
to convert from the frequency domain to the time or 
distance domain by defining
\beq
b(t) = \int_{\Omega-B}^{\Omega+B} dk b(k)e^{-i(k-\Omega)t}.
\eeq
These obey 
\beq
[b(t),b\dg(t')] = \delta(t-t'), \label{comrel2}
\eeq
where $b\dg(t) \equiv [b(t)]\dg$, and the $\delta$ function 
is actually a narrow function with width of 
order $B^{-1}$. The operator 
$b\dg(t)b(t)$ can be interpreted as the photon flux operator.

Defining continuum 
quadrature operators
\beq
X(t)= b(t)+b\dg(t) \;,\;\;
Y(t) = -ib(t)+ib\dg(t),
\eeq
one obtains 
\beq
[X(t),Y(t')] = 2i\delta(t-t').
\eeq
The singularity in the associated uncertainty relations can be 
avoided by quantifying the uncertainty by the {\em spectrum} 
\cite{WalMil94}
\beq 
S^{X}(\omega) = \ip{\tilde{X}(\omega)X(0)}_{\rm ss} - 
\ip{\tilde{X}(\omega)}_{\rm ss}\ip{X(0)}_{\rm ss}.
\eeq
Then one can derive the finite uncertainty relations \cite{Sha87}
\beq \label{uncrel1}
S^{X}(\omega)S^{Y}(\omega) \geq 1.
\eeq

For very high frequencies the spectra always go to unity. This 
represents the shot-noise or vacuum noise level. However for finite 
frequencies it is possible to have for example $S_{X}(\omega)<1$. This 
indicates squeezing of the $X$ quadrature. The uncertainty in the 
conjugate quadrature would of course be increased. 

The quadrature operators $X(t),Y(t)$ can be directly measured (one at 
a time) using homodyne detection \cite{YueSha80}. 
For detection of efficiency $\epsilon \leq 1$ the photocurrent 
$I_{\rm hom}^{X}$ has the same statistics as (and therefore can be 
represented by) the operator
\beq
I_{\rm hom}^{X}(t) = \rt{\epsilon}X(t) + \rt{1-\epsilon}\xi_{X}(t),
\eeq
where $\xi_{X}(t)$ is a Gaussian white noise term. Here the normalization 
has been chosen so that the  photocurrent
spectrum
\beq
S^{X}_{\rm hom}(\omega) = \epsilon S^{X}(\omega) + (1-\epsilon),
\eeq
remains equal to unity (the shot noise limit) at high 
frequencies, where $S^{X}(\omega)=1$.

\subsection{An Atom in a Squeezed Bath}
\label{atomsec1}

Now consider the situation where a two-level atom is immersed in a beam of 
squeezed light with annihilation operator $b_{0}(t)$. 
If the degree of mode matching of the squeezed light 
to the atom's dipole radiation mode is $\eta$, then the dipole 
coupling Hamiltonian in the rotating-wave approximation is 
\beq \label{vac}
H_{0}(t) = -i[\rt\eta b_{0}(t) + 
\rt{1-\eta}\nu(t)]\sigma\dg(t) + {\rm H.c.} 
\eeq
Here $\sigma$ is the atomic lowering operator and the atomic 
linewidth has been set to unity. The operator $b_0(t)$ 
represents the squeezed field with spectra $S^{X}_0(\omega)$ and 
$S^{Y}_0(\omega)$, and $\nu(t)$ represents the vacuum 
field interacting with the atom, satisfying 
$\ip{\nu(t)\nu\dg(t')}=\delta(t-t')$. 

Now if the quadrature spectra of the 
squeezed light are much broader than the atomic linewidth then one 
can make the white noise approximation that they are constant. For 
 minimum-uncertainty squeezing, one has
\beq
S^{X}_{0}(\omega) = L = 1/S^{Y}_{0}(\omega),
\eeq
and the atom will obey the master equation \cite{Gar86}
\beq
\dot\rho = (1-\eta){\cal D}[\sigma]\rho + \frac{\eta}{4L}
{\cal D}[(L+1)\sigma-(L-1)\sigma\dg]\rho,
\eeq
where ${\cal D}[A]B \equiv A\dg B A - \half \{A\dg A ,B\}$ as usual.  
This leads to the following atomic dynamics:
\bqa
{\rm Tr}[\dot\rho \sigma_{x}] &=& -\gamma_{x}{\rm Tr}[\rho 
\sigma_{x}], \label{onex}\\
{\rm Tr}[\dot\rho \sigma_{y}] &=& -\gamma_{y}{\rm Tr}[\rho 
\sigma_{y}],\\
{\rm Tr}[\dot\rho \sigma_{z}] &=& -\gamma_{z}{\rm Tr}[\rho 
\sigma_{z}]-C \label{threez},
\eqa
where
\bqa
\gamma_{x} &=& \half\left[ (1-\eta)+\eta L\right] \label{gx2},\\
\gamma_{y} &=& \half\left[ (1-\eta)+\eta L^{-1}\right], \label{gy2} \\
\gamma_{z} &=& \gamma_{x}+\gamma_{y} \;,\;\;C=1.
\eqa

Note that for $L< 1$  the decay rate of the $x$ 
component of the atomic dipole is reduced 
below the vacuum level of $\half$, 
while the decay rate of the other component is 
increased.  The reduction or increase in the decay rates are directly 
attributable to the 
reduction or increase in the fluctuations of the respective 
quadrature of the input continuum field. The prediction of this 
effect by Gardiner \cite{Gar86} 
began the study of quantum spectroscopy (that is, the interaction of 
nonclassical light with matter) \cite{FicDru97}. 

For sufficiently large $\eta$ this effect would be 
easily detectable experimentally 
in the fluorescence power spectrum 
of the atom (into the vacuum modes):
\bqa \label{genps}
P(\omega) &=& \frac{1-\eta}{2\pi}\ip{\tilde{\sigma}\dg(-\omega)
\sigma(0)}_{\rm ss} \\
&=& \frac{(1-\eta)(\gamma_{z}-C)}{8\pi\gamma_{z}}\left[ 
\frac{\gamma_{x}}{\gamma_{x}^{2}+\omega^{2}} + 
\frac{\gamma_{y}}{\gamma_{y}^{2}+\omega^{2}}\right].
\eqa
For $L<1$ the spectrum consists of two Lorentzians, one with a 
sub-natural linewidth and one with a super-natural linewidth. 
The overall linewidth (defined as the full-width at half-maximum height)
is reduced.

This line-narrowing is only noticeable if the degree of mode matching $\eta$ of 
the squeezed light to the 
atom is significant, which is hard to do with a squeezed beam.
One way around this is to make the atom strongly coupled to a 
microcavity, which can be driven by a squeezed beam. The microcavity 
enhances the atomic decay rate, but a squeezed input should suppress 
this enhancement in one quadrature. Unfortunately, 
experiments to date have failed to see this suppression,
due to imperfections of various kinds \cite{Tur98}.

\section{Theory of Squashed States}

\subsection{Generation of Squashed States}

Consider the feedback loop shown in Fig.~1. The field entering the 
modulator is $b_{0}(t)= \half[X_{0}(t)+iY_{0}(t)]$. The quadrature 
operators are assumed to have  independent statistics defined by the 
spectra $S^{X}_{0}(\omega)$, $S^{Y}_{0}(\omega)$.
The modulator simply adds a coherent amplitude to this field. There 
are various ways of achieving this, one of which is discussed in 
Ref.~\cite{Wis98b}. The field exiting is in any case given by
\beq \label{b1}
b_{1}(t) = \half\left[X_{0}(t)+iY_{0}(t) + \chi(t) + i\upsilon(t)\right],
\eeq
where $\chi(t),\upsilon(t)$ are real functions of time. This field now 
enters a homodyne detection device, set up so as to measure the $X$ 
quadrature. If the efficiency of the measurement is $\epsilon_{X}$ 
then the photocurrent is given by
\beq \label{ihom}
I_{\rm hom}^{X}(t) = \rt{\epsilon_{X}}
\left[X_{0}(t)+ \chi(t) \right] + \rt{1-\epsilon_{X}}\xi_{X}(t).
\eeq

Now, through the feedback loop, this current may determine the 
classical field amplitudes $\chi,\upsilon$. Obviously in the case 
of measuring the $X$ quadrature, 
the only interesting results will come from controlling $\chi(t)$, 
and we set
\beq
\chi(t) = \int_{0}^{\infty} g_{X} h(s) 
I_{\rm hom}^{X}(t-\tau-s)/\rt{\epsilon_{X}}ds.
\eeq
Here $\tau$ is the minimum delay in the feedback loop, $h(s)$ is 
the feedback loop response function normalized to  
$\int_{0}^{\infty}h(s)ds = 1$ and $g_{X}$ is the round-loop gain \cite{Ste90}. 
Taking the fourier transform of this expression and substituting 
into Eqs.~(\ref{b1}) and (\ref{ihom}) yields
\beq \label{b1p}
\tilde{b}_{1}(\omega) = \half\left[
\frac{\tilde{X}_{0}(\omega)+
\rt{\theta_{X}}g_{X} e^{i\omega\tau}\tilde{h}(\omega)
\tilde{\xi}_{X}(\omega)}
{1-g_{X} e^{i\omega\tau}\tilde{h}(\omega)}+i\tilde{Y}_{0}(\omega)  
+ i\upsilon(t)\right].
\eeq
where
\beq
\theta_{X} \equiv \epsilon_{X}^{-1}-1.
\eeq

The $X$ quadrature spectrum of this light is
\beq
S_{1}^{X}(\omega) = 
\frac{S_{0}^{X}(\omega)+\theta_{X}g_{X}^{2}|\tilde{h}(\omega)|^{2}}
{|1-g_{X} e^{i\omega\tau}\tilde{h}(\omega)|^{2}}.
\eeq
 Say we are only interested in frequencies well 
inside the bandwidth of $\tilde{h}(\omega)$, much less than 
$\tau^{-1}$, and much less than the bandwidth of $S_{0}^{X}(\omega)$. 
Then we can replace $e^{i\omega\tau}\tilde{h}(\omega)$ by unity and 
$S_{0}^{X}(\omega)$ by a constant $L$. This gives
\beq \label{conspec}
S_{1}^{X} = \frac{L+g_{X}^{2}\theta_{X}}{(1-g_{X})^{2}} \geq 
\frac{L}{1+L/\theta_{X}},
\eeq
where the minimum is achieved for negative feedback $g_{X} = -L/\theta_{X}$. 

Evidently this minimum is less than $L$. This means that even starting with 
shot-noise limited light ($L=1$) it is possible to produce 
sub-shot-noise light. However, it is important to note that this is not 
squeezed light in the ordinary sense. For example, it is impossible to 
remove any of the squeezed light by putting a beam splitter in the 
path of the in-loop beam. Under the above conditions 
the resulting out-of-loop beam actually has a noise level above the shot noise 
\cite{WalJak85,YamImoMac86,Sha87,Tau95}. 
Nevertheless, the fluctuations of the in-loop light 
do produce genuine physical effects in other circumstances, as investigated 
in Sec.~4.

\subsection{Violation of the Uncertainty Relations}

A 
curious point in the apparent squeezing of the $X$ quadrature is that 
the feedback has no effect on the $Y$ 
quadrature of $b_{1}$. From \erf{b1p}, $S_{1}^{Y} = L^{-1}$, assuming 
a minimum uncertainty input $b_{0}$ and $\upsilon=0$. Thus the 
uncertainty relation (\ref{uncrel1}) is violated for this in-loop 
light. For this reason, the in-loop light exhibiting sub-shot-noise 
fluctuations has been called ``squashed light'' \cite{Buc99}. By 
feedback, the noise in one quadrature can be squashed (reduced), but 
there is no ``squeezing'' of phase-space area into an increased noise 
in the other quadrature.

The reason that the uncertainty relation (\ref{uncrel1}) is violated 
is that the commutation relations (\ref{comrel2}) are no longer valid 
for time differences $|t-t'| > \tau$, the minimum feedback loop delay. 
This is a direct consequence of 
the feedback loop. It is important to realize that the parts of the 
in-loop field separated in time by greater than $\tau$ never actually 
exist together. That is because the propagation time from the 
modulator to the detector is necessarily less than $\tau$. Thus the 
fundamental commutation relations \cite{Gar91} between parts of the 
field at different points in {\em space} at the same time are never violated.

For freely propagating fields there is no real distinction between space and time 
separations, but for an in-loop field it is a crucial distinction. 
The temporal anticorrelations in the in-loop squeezed 
light only exist for time separations greater than $\tau$, and hence 
greater than the time for which any part of the in-loop light exists. There is 
never any anticorrelation between parts of the in-loop field in 
existence at any given time. By contrast, conventional squeezed light 
can propagate for an arbitrarily long time before detection, so the 
anticorrelations are between parts of the field which can exist 
simultaneously (even if they may not actually do so in a given 
experiment).  

\subsection{Simultaneous Squashing in Both Quadratures}

It is interesting now to consider feedback in both the $X$ and $Y$ 
quadratures. Obviously one cannot simultaneously measure both of these 
quadratures with unit efficiency, but one can measure $Y$ with 
efficiency $\epsilon_{Y} \leq 1- \epsilon_{X}$. Carrying through the same sort 
of analysis as above shows that the $Y$ quadrature spectrum can be 
simultaneously reduced to 
\beq
S_{1}^{Y} = \frac{L^{-1}}{1+L^{-1}/\theta_{Y}}.
\eeq
For the special case of $L=1$ (a vacuum input), one finds
\beq
S_{1}^{Y} + S_{1}^{X} = 2 - \epsilon_{X}  - \epsilon_{Y} \geq 1.
\eeq
For feedback based on heterodyne detection (which is equivalent to 
homodyne detection on both quadratures with equal efficiency), one can 
have $S_{1}^{Y} = S_{1}^{X} = 1 - \half \epsilon$, which goes to 
one half in the limit of perfect detectors.

\subsection{Simultaneous Squeezing and Squashing}
\label{eezash}
In the general case of $L \neq 1$, the sum of the quadrature spectra 
need not even be greater than one. 
Rather, for fixed $L<1$ and fixed $\epsilon = 
\epsilon_{X}+\epsilon_{Y}$ one finds, for $\epsilon_{X}=0$ and 
$\epsilon_{Y} = \epsilon$ 
\beq
S_{1}^{Y} + S_{1}^{X} = L + 
\frac{\epsilon^{-1}-1}{L(\epsilon^{-1}-1) + 1} \geq 0,
\eeq
where the limit of zero noise in both quadratures is approached for 
$L\to 0$ and $\epsilon \to 1$. For example, with experimentally 
realisable parameters of 6dB squeezing \cite{Tur98}
 and detection efficiency $\epsilon = 0.95$ \cite{Sch96}, one could obtain 
$S_{1}^{X} = 0.25$ and $S_{1}^{Y} = 0.05$, giving 
$S_{1}^{Y}+S_{1}^{X} = 0.30$, compared to the limit of 
$2$ implied by the uncertainty relation (\ref{uncrel1}).

\section{Application to Quantum Spectroscopy}

Since the quadrature spectra calculated above apply to an in-loop 
field, which cannot be extracted using a beam splitter, it might seem 
that they have no physical significance. However, this is not the 
case. As I showed recently \cite{Wis98b}, placing a two-level atom in 
an squashed bath leads to linewidth narrowing of one atomic dipole
quadrature, entirely analogous to 
that produced by a squeezed bath. The master equation is not identical, 
however, because the 
non-squashed quadrature is still shot-noise limited, so there is no 
line broadening of the other atomic dipole quadrature.

In this work I generalize the results of Ref.~\cite{Wis98b} to include 
light which is simultaneously squashed in both quadratures, or 
simultaneously squeezed and squashed. If the atom is coupled to 
the in-loop beam $b_{1}(t)$ with mode-matching $\eta$ then the atomic 
Hamiltonian is  
\beq \label{totalHam}
H(t)= -i[\rt\eta b_{1}(t) + 
\rt{1-\eta}\nu(t)]\sigma\dg(t) + {\rm H.c.}
\eeq
Following the methods of Ref.~\cite{Wis98b}, the expression for 
$b_{1}$ is modified from (\ref{b1p}) by the addition of the atom's 
radiated field in the direction of the beam, 
$\rt{\eta}\tilde{\sigma_{x}}(\omega)$, to the input operator 
$\tilde{X}_{0}(\omega)$. 
Thus the total Hamiltonian can be written
\beq
H(t)= H_{\rm fb}(t) + H_{0}(t),
\eeq
where $H_{0}(t)$ is as given in \erf{vac} and the Hamiltonian 
due to the feedback is
\bqa 
H_{\rm fb}(t) &=&  \lambda_{X} \half \sigma_{y}(t)\left\{ \sigma_{x}(t^{-}) + 
[X_{0}(t^{-})+\rt{\theta_{X}}\xi_{X}(t^{-})]\sqrt{\eta}\right\} \nn\\
&& \,+  \lambda_{Y} \half\sigma_{x}(t)\left\{ \sigma_{y}(t^{-}) + 
[Y_{0}(t^{-})+\rt{\theta_{Y}}\xi_{Y}(t^{-})]\sqrt{\eta}\right\} 
.\label{Hfb}
\eqa
The feedback parameters are defined as  
\beq
\lambda_{X} = \frac{g_{X}\eta}{1-g_{X}}\;;\;\;
\lambda_{Y} = \frac{g_{Y}\eta}{1-g_{Y}}.
\eeq
for the feedback of the homodyne current $I_{\rm 
hom}^{Y}$ are defined analogously to those from the feedback of 
$I_{\rm hom}^{X}$.

In \erf{Hfb} the limit of broad-band feedback has been taken, with 
$\tilde{h}(\omega)e^{i\omega\tau}$ in \erf{b1p} set to unity. This Markov 
approximation is justified
provided the bandwidth of the feedback is very large compared to the 
characteristic rates of response of the system \cite{Wis94a}. 
In the present context the rate of atomic decay is unity so we 
require, for instance, $\tau \ll 1$. For a typical electro-optic 
feedback loop with a bandwidth in the MHz range, the atom would have 
to have be metastable to satisfy this inequality. 
The precise requirements for the validity of the 
Markov approximation will be investigated in a future publication. 
Of course even in the broad-band 
limit the feedback from the measurement of a particular part of the 
field must act after that part of the field 
has interacted with the atom. This is the reason for the use of the time argument 
$t^{-}$ rather than $t$ in \erf{Hfb}.

Now to describe the evolution generated by the total Hamiltonian 
(\ref{totalHam}), the  theory of homodyne detection and feedback in 
the presence of white noise is required. This was first detailed 
in Ref.~\cite{WisMil94b}, generalizing the earlier work in 
Refs.~\cite{WisMil93b} and \cite{Wis94a}. 
The basic equation is
\beq \label{cke}
d\rho(t) = \ip{\exp[-iH_{\rm fb}(t)dt]\exp[-iH_{0}(t)dt]
\rho(t)\exp[iH_{0}(t)dt]\exp[iH_{\rm fb}(t)dt]}-\rho(t).
\eeq
Here the ordering of the unitary operators has been chosen such that 
the time delay of the feedback has been taken into 
account and one can replace $t^{-}$ in \erf{Hfb} by $t$. 

In \erf{cke} the 
expectation value indicates an average over the bath operators 
$b_{0}(t)$  and $\nu(t)$, and 
the detector noise terms $\xi_{X}(t),\xi_{Y}(t)$. This is effected by 
making replacements such as
\beq
[X_{0}(t)dt]^{2} \to Ldt\;;\;\; \nu(t)dt\, \nu\dg(t)dt \to dt.
\eeq
The result is 
\bqa
\dot{\rho} &=& (1-\eta){\cal D}[\sigma]\rho + \frac{\eta}{4L}
{\cal D}[(L+1)\sigma-(L-1)\sigma\dg]\rho \nn \\
&& -i{\lambda_{X}}\left[ 
\half\sigma_{y},\half\{(L+1)\sigma-(L-1)\sigma\dg\}
\rho+\rho\half\{(L+1)\sigma\dg-(L-1)\sigma\}\right] \nn \\
&& + i{\lambda_{Y}}\left[ 
\half\sigma_{x},-i\half\{(L^{-1}+1)\sigma+(L^{-1}-1)\sigma\dg\}
\rho+i\rho\half\{(L^{-1}+1)\sigma\dg-(L^{-1}-1)\sigma\}\right]
\nn \\
&& + \frac{\lambda_{X}^{2}(L+\theta_{X})}{\eta}{\cal 
D}\left[\frac{\sigma_{y}}{2}\right]\rho 
+ \frac{\lambda_{Y}^{2}(L^{-1}+\theta_{Y})}{\eta}
{\cal D}\left[\frac{\sigma_{x}}{2}\right]\rho.
\eqa
This again produces the atomic dynamics of Eqs.~(\ref{onex})--(\ref{threez}), 
but with
\bqa
\gamma_{x} &=& \half\left[ (1-\eta)+\eta 
L(1+\lambda_{X}/\eta)^{2}+\lambda_{X}^{2}\theta_{X}/\eta\right] = \half
\left[ (1-\eta)+\eta S_{X}\right],\\
\gamma_{y} &=& \half\left[ (1-\eta)+\eta 
L^{-1}(1+\lambda_{Y}/\eta)^{2}+\lambda_{Y}^{2}\theta_{Y}/\eta\right] = \half
\left[ (1-\eta)+\eta S_{Y}\right], \\
\gamma_{z} &=& \gamma_{x}+\gamma_{y} \;,\;\;C=1+\lambda_{X}+\lambda_{Y}.
\eqa

In the above equations, the introduction of the spectrum $S^{X}$ is based on 
\erf{conspec}, with the identification $\lambda_{X} = \eta g_{X} /(1-g_{X})$,
and similarly for $Y$. Note that the expressions for $\gamma_{x}$ 
and $\gamma_{y}$ depend upon the quadrature spectra (in the absence 
of the atom) in precisely the 
same way as those for pure squeezed (not squashed) light in 
Eqs.~(\ref{gx2}) and (\ref{gy2}). This suggests that the natural 
explanation for the change in the decay rates is again the reduced 
fluctuations of the input light. It seems that squashed fluctuations 
are much the same as squeezed fluctuations as far as the atom is 
concerned. The one difference is that the constant $C$ in the equation 
of motion for $\ip{\sigma_{z}}$ is also altered  by the feedback.

Consider now the case as in Sec.~\ref{eezash} 
where $L<1$, $\epsilon_{X}=0$ and 
$\epsilon_{Y}=\epsilon$. Then choosing 
$\lambda_{Y}=-\eta/(1+L\theta)$, so as to minimize 
the in-loop $Y$ quadrature spectrum, gives
\bqa
\gamma_{x} &=& \half\left[ (1-\eta)+\eta L\right] ,\\
\gamma_{y} &=& \half\left[ (1-\eta)+\eta{\theta}/({1+L\theta})\right], \\
\gamma_{z} &=& \gamma_{x}+\gamma_{y} \;,\;\;C=1 - {\eta}/({1+\theta 
L}),
\eqa
where $\theta = \epsilon^{-1}-1$ as usual. Choosing $\epsilon=0.95$ 
and $L\approx 0.25$, as before, gives $\gamma_{z} \approx 1 - 0.85 \eta$. 
That is, in the limit of $\eta \to 1$, the rate of decay of the atomic 
population would be slowed by $85\%$. In the ideal limit 
of $L \to 0$, and $\epsilon,\eta \to 1$, the atom would be frozen in 
its initial state and would not decay at all.

\section{Conclusion}

In this work I have presented for the first time the theory for 
a new class of in-loop light, namely light which may be both  
squashed (in either or both quadratures) and squeezed. 
Squeezing here refers to conventional quantum noise 
reduction, whereas squashing refers to the noise reduction produced by 
the feedback loop. Even without a squeezed input it is possible to reduce 
the noise in {\em both} quadratures of the in-loop field below the 
shot-noise limit. With a squeezed input it is possible, in principle, 
to reduce the noise in both quadratures to zero (by squeezing one and 
squashing the other). 

I next derived the effect of this arbitrarily squeezed and squashed 
light on an in-loop atom. The calculated in-loop spectra 
precisely reflect the noise to which the atom responds, provided the 
bandwidth of the squeezing and the bandwidth of the feedback are much 
greater than the atomic linewidth. As  
the quantum fluctuations seen by the atom are reduced, the decay rates for the 
quadratures of the atom's dipole are reduced. In the limit that the 
atom is coupled only to in-loop 
light which is perfectly squeezed in one quadrature and perfectly 
squashed in the other, the atomic decay rates vanish and the atom's 
dynamics are frozen.

As discussed above, the main experimental difficulty with seeing 
squeezing-induced line-narrowing is related to efficiently 
coupling the squeezed light onto the atom. Using squashed light 
rather than squeezed light would not overcome this difficulty. 
However, highly squashed light  should be easier to generate than 
highly squeezed light, because it is limited only by the homodyne detector 
efficiency. Also, it can be produced at any frequency for which a 
coherent source and the 
appropriate electro-optic equipment is available. These factors, plus 
the intriguing possibility of observing simultaneous linewidth 
narrowing on both atomic quadratures, suggest an important role
for squashed light in experimental quantum spectroscopy.

\begin{figure}
	\vspace{6cm}
	\begin{center}
		\special{illustration 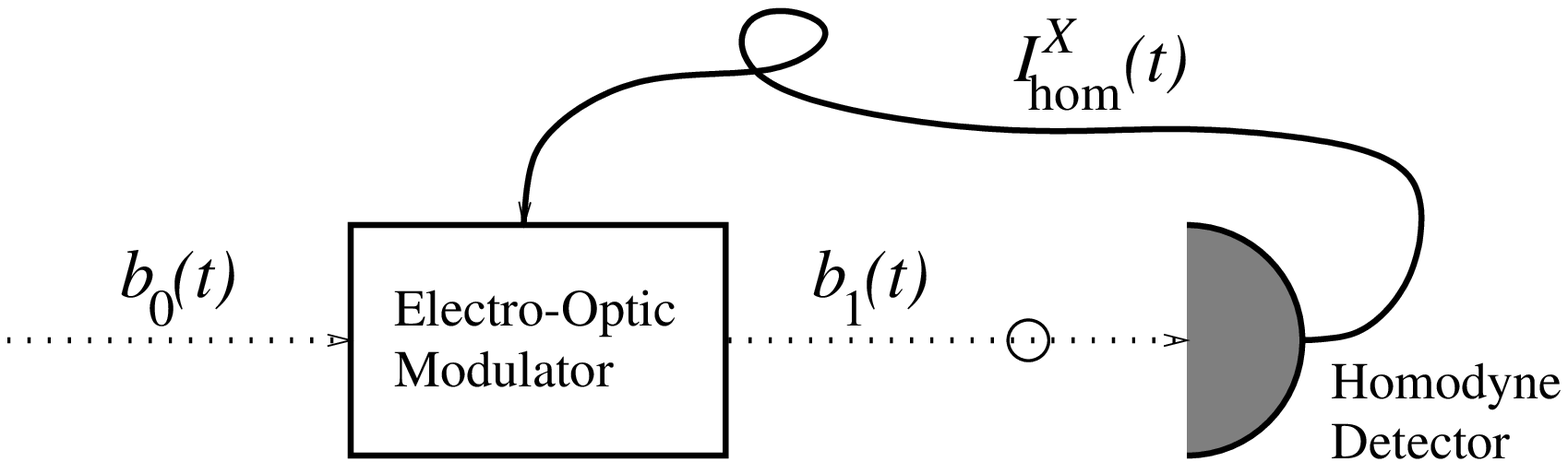 scaled 0.85}
	\end{center}
	\vspace{-0.5cm}
	\caption{Simplified diagram of the production of squashed light. 
	The circle in the path of the in-loop beam is where an in-loop 
	atom could be positioned. 
	}
	\protect\label{fig1}
\end{figure}

\end{document}